\documentclass[10pt,twocolumn,letterpaper]{article}

\usepackage{iccv}
\usepackage{times}
\usepackage{epsfig}
\usepackage{graphicx}
\usepackage{amsmath}
\usepackage{amssymb}

\usepackage[export]{adjustbox}
\usepackage[caption=false]{subfig}
\usepackage[usenames,dvipsnames]{color}
\usepackage[pagebackref=true,breaklinks=true,colorlinks,bookmarks=false]{hyperref}
\usepackage{booktabs}
\usepackage{multirow}
\usepackage{array}
\newcommand{\ws}{\! }  

\newcolumntype{x}{>{\centering\arraybackslash}p{0.5cm}}
\newcolumntype{X}{>{\centering\arraybackslash}p{2.1cm}}
\newcommand{\anonymized}[1]{#1}

\iccvfinalcopy  
\ificcvfinal\fi  

\def\iccvPaperID{41}  

\begin{document}

\title{Improving Robustness of Deep Learning Based Knee MRI Segmentation: Mixup and Adversarial Domain Adaptation}

\author{
    Egor Panfilov$^{1}$ \\
\and
    Aleksei Tiulpin$^{1,2}$ \\
\and
    Stefan Klein$^{3}$ \\
\and
    Miika T. Nieminen$^{1,2}$ \\
\and
    Simo Saarakkala$^{1,2}$ \\
    \small$^1$University of Oulu, Oulu, Finland \hspace{1pt} $^2$Oulu University Hospital, Oulu, Finland \hspace{1pt} $^3$Erasmus MC, Rotterdam, The Netherlands\\
    \small\texttt{egor.panfilov@oulu.fi}\\
}

\maketitle
\ificcvfinal\thispagestyle{empty}\fi  

\begin{abstract}
Degeneration of articular cartilage (AC) is actively studied in knee osteoarthritis (OA) research via magnetic resonance imaging (MRI). Segmentation of AC tissues from MRI data is an essential step in quantification of their damage. Deep learning (DL) based methods have shown potential in this realm and are the current state-of-the-art, however, their robustness to heterogeneity of MRI acquisition settings remains an open problem. In this study, we investigated two modern regularization techniques --  mixup and adversarial unsupervised domain adaptation (UDA) -- to improve the robustness of DL-based knee cartilage segmentation to new MRI acquisition settings. Our validation setup included two datasets produced by different MRI scanners and using distinct data acquisition protocols. We assessed the robustness of automatic segmentation by comparing mixup and UDA approaches to a strong baseline method at different OA severity stages and, additionally, in relation to anatomical locations.
Our results showed that for moderate changes in knee MRI data acquisition settings both approaches may provide notable improvements in the robustness, which are consistent for all stages of the disease and affect the clinically important areas of the knee joint. However, mixup may be considered as a recommended approach, since it is more computationally efficient and does not require additional data from the target acquisition setup.
\end{abstract}
\section{Introduction}
Knee osteoarthritis (OA) is the most common musculoskeletal disease in the world. OA is poorly understood and no disease-modifying treatment currently exists for it~\cite{van2019osteoarthritis}.
Magnetic resonance imaging (MRI) methods are commonly used to clinically study the structural changes within the knee joint and, specifically, in articular cartilage~\cite{nieminen2018osteoarthritis}. A variety of MRI acquisition protocols has been introduced, each tailored to visualize specific tissues of interest or measure particular tissue properties~\cite{abdulaal20173t,altahawi20183d}. Moreover, there is a large number of MR scanner models available on the market, with major differences in hardware and reconstruction software. As a result, MR images qualitatively vary from institution to institution, from study to study, and from dataset to dataset.

Since OA is a long-term and complex disease, large longitudinal studies have been carried out to investigate onset and progression of OA. Currently, one of the major areas of interest is assessment of compositional and morphological changes in articular cartilage tissues~\cite{emery2019establishing}. In order to perform these analyses from MRI, the tissues need to be segmented. However, manual delineation of cartilage tissues is time-consuming, prone to high intra- and inter-rater variability~\cite{stammberger1999interobserver}, and challenging due to the large size of datasets and the aforementioned issues related to data heterogeneity. Consequently, there is a clear need for automatic methods for knee cartilage segmentation, which are accurate and robust to variations in data acquisition setting.

Recently, in OA and other fields of medicine, deep learning (DL) methods have become the new state-of-the-art in computer-aided diagnosis~\cite{tiulpin2019automatic,tiulpin2019multimodal,wang2019grey,tiulpin2018automatic,esteva2017dermatologist,esteva2019guide}. Latest advances in automatic segmentation methods, in particular DL-based, have demonstrated promising results in knee tissue segmentation~\cite{norman2018relaxometry,tack2018menisci,ambellan2019bonecartilage,desai2019considerations,tack2019accurate,chaudhari2019utility}. Such methods produce accurate and consistent results, but they often lack evaluation on independent datasets and, therefore, are potentially prone to large variations in the input data characteristics. The issue originates from the fact that supervised DL-based algorithms, when trained on medical imaging datasets that are often limited in size and diversity, tend to incorporate dataset bias and fail to generalize to new domains~\cite{karani2018lifelong}.

In this paper, we focus on regularization of DL-based methods for knee cartilage segmentation from MRI, and investigate two state-of-the-art approaches to improve the generalization to new data. The contributions of this study are the following:

\begin{itemize}
    \item We introduce an efficient and accurate DL-based baseline method for knee cartilage segmentation that performs comparably or improves on the previous state-of-the-art.
    \item We investigate the use of an end-to-end unsupervised domain adaptation (UDA) approach for knee MRI segmentation, and show how both labelled and unlabelled data can be leveraged within the same segmentation framework.
    \item We explore the use of data augmentation via mixup in the considered semantic segmentation problem, and report its effectiveness in multiple setups.
    \item We validate the baseline method and its modifications with mixup and UDA on an independent test set and demonstrate the improvements in model robustness. We also provide a detailed analysis of the results and examine the performance of the methods in relation to the anatomical locations that are the most clinically relevant (e.g. weight bearing areas of the knee joint).
    \item Finally, we make our source code and the pre-trained models publicly available.
\end{itemize}

\section{Related Work}
Due to the wide adoption of MR imaging methods, semi-automatic and automatic knee cartilage segmentation from MRI has been studied already for several decades, with more focus recently on purely automatic methods~\cite{ryzhkov2015knee,kumar2018knee}. However, despite the availability of large imaging cohorts, such as Osteoarthritis Initiative (OAI)~\cite{peterfy2008osteoarthritis}, large-scale analysis of such data in scope of OA research remains extremely challenging due to the lack of annotations. Same applies to the datasets from numerous hospitals, which are typically less standardized and the annotations are even more sparse and of lower quality.

Several recent studies~\cite{norman2018relaxometry,liu2018deformable,ambellan2019bonecartilage,desai2019considerations} have shown that specifically DL-based approaches for knee cartilage and meniscus segmentation can achieve accuracy close to the human level and superior to the conventional atlas-based methods~\cite{dam2015segmentation}.
However, no validation of those DL-based methods on independent datasets acquired in various hospitals has yet been conducted. Therefore, the general applicability of all the previously published DL-based cartilage segmentation methods remains unclear.

To tackle the robustness-related issues in modern deep neural networks, a wide range of techniques of different complexity has  been proposed~\cite{kukavcka2017regularization}. Their effectiveness in the specific tasks and domains, however, is still to be practically investigated.

One of the recent effective techniques to improve model generalization and reduce memorization of the training data was mixup~\cite{zhang2018mixup}. The idea of mixup was to use a convex combination of the inputs and the targets to augment the training data with such interpolated examples. Mixup has been applied in several image classification problems and has shown to notably reduce the overfitting and stabilize the convergence of models~\cite{yaguchi2019mixfeat,guo2019mixup}. Nonetheless, the applicability of the technique and its performance in semantic segmentation problems remains unclarified, and very few studies investigated this topic~\cite{eaton2018improving,chaitanya2019semi}. Our goal was to adapt the technique to knee cartilage segmentation problem and evaluate its performance in different settings.

Another approach that has attracted great interest during the recent years is domain adaptation (DA)~\cite{daume2006domain,chopra2013dlid}. A great number of the popular DA techniques is based on the following hypothesis: in order to have a good generalization for any machine learning method, the representations of data samples (including the samples from different domains or datasets) have to share a large common subspace or be somewhat aligned. Ganin~\emph{et al.}~\cite{ganin2016domain} and Tzeng~\emph{et al.}~\cite{tzeng2014deep} were among the first to discover how this alignment can be applied to DL-based models, and moreover, how to perform it in a semi-supervised way, such that both labelled and unlabelled data from different domains can be incorporated into the training process. The framework was called Unsupervised Domain Adaptation (UDA) and its potential for the development of robust models has been shown in various applications.

In medical imaging, UDA has been studied in several fields for which multiple diverse datasets are publicly available: brain MRI~\cite{karani2018lifelong,chaitanya2019taskdriven}, chest X-ray~\cite{chen2018semanticaware}, cardiac MRI-CT~\cite{dou2018crossmodality} and others. However, very few studies investigated the use of UDA techniques in knee MRI domain and, more specifically, knee cartilage segmentation. Joint multitask learning of deep segmentation and registration networks was suggested by Xu~\emph{et al.}~\cite{xu2019deepatlas}. Liu~\cite{liu2019susan} proposed to train joint segmentation and cycle-consistent image-to-image translation between the labelled and unlabelled domains. Both of the approaches, however, are rather complicated and depend on the performance of an auxiliary task -- registration or image-to-image translation respectively.

In this work, we explored UDA via cross-domain alignment of deep representation spaces. The chosen method has only moderate computational costs, can be easily scaled to larger number of domains and extended with other regularization techniques.
\section{Materials and Methods}
\subsection{Problem Statement}
\begin{figure}[ht!]
    \centering
    \includegraphics[width=1.0\linewidth]{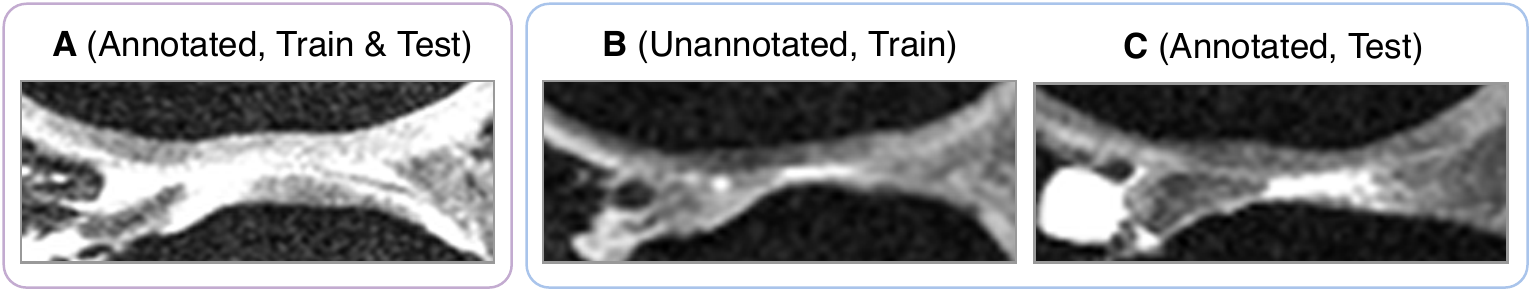}
    \caption{Examples of MRI images from Datasets A, B, and C. Here, we show only the tibiofemoral areas, which enclose femoral cartilage, tibial cartilage, and menisci.}
    \label{fig:datasets}
\end{figure}

Our goal was to assess how different regularization techniques, namely, mixup and UDA, perform in combination with a strong baseline method yielding state-of-the-art results in knee cartilage segmentation task. To assess the generalization, we validated all our methods on independent data. For UDA, we assumed that the unlabelled data from a dataset similar to the test dataset is available during training. In total, our setup included three datasets: two datasets from different MRI scanners and data acquisition protocols (Datasets A and B), and a third dataset (Dataset C) acquired similarly to Dataset B, but in an independent clinical study (see Figure~\ref{fig:datasets}). Train subset of Dataset A and whole Dataset B were used for development of the approaches. Test subset of Dataset A and whole Dataset C were used for evaluation purposes.

Let $\textbf{X}^a$ be a mini-batch of image samples from the annotated dataset, $\textbf{Y}^a$ - corresponding reference annotations, $\textbf{X}^b$ - a mini-batch of image samples from the unannotated dataset. $S$ is a model that takes as input a mini-batch of images $\textbf{X}$ (either $\textbf{X}^a$ or $\textbf{X}^b$) and produces the segmentation masks $\hat{\textbf{Y}}$. In all the experiments the models were trained to perform the segmentation slice-wise (i.e. in 2D).
\subsection{Baseline Method}
Our baseline approach (Figure~\ref{fig:scheme_bl}) was based on U-Net~\cite{ronneberger2015unet}. Similarly to~\cite{tiulpin2019tidemark} we used $24$ filters in the first convolutional block and doubled the number of filters at each depth level. The total model depth was set to $6$. In the expanding path we used bilinear upsampling instead of $2\times 2$ up-convolutions. As a results of an extensive experimental search we found that such model parameters yielded the best performance in the considered task. The network was trained to produce $5$ mutually exclusive segmentation masks: no cartilage, femoral cartilage (FC), tibial cartilage (TC), patellar cartilage (PC), and menisci (M). For training we used multi-class cross-entropy ($\textrm{MCE}$) loss, which was calculated between the randomly sampled masks $\textbf{Y}^{a}$ and the model predictions $\hat{\textbf{Y}}$ produced from the corresponding $\textbf{X}^{a}$.
\subsection{Regularization Techniques}
\begin{figure}[ht!]
    \begin{center}
    \subfloat[\label{fig:scheme_bl}]{
      \includegraphics[trim=12 5 14 0,clip,width=1.0\linewidth]{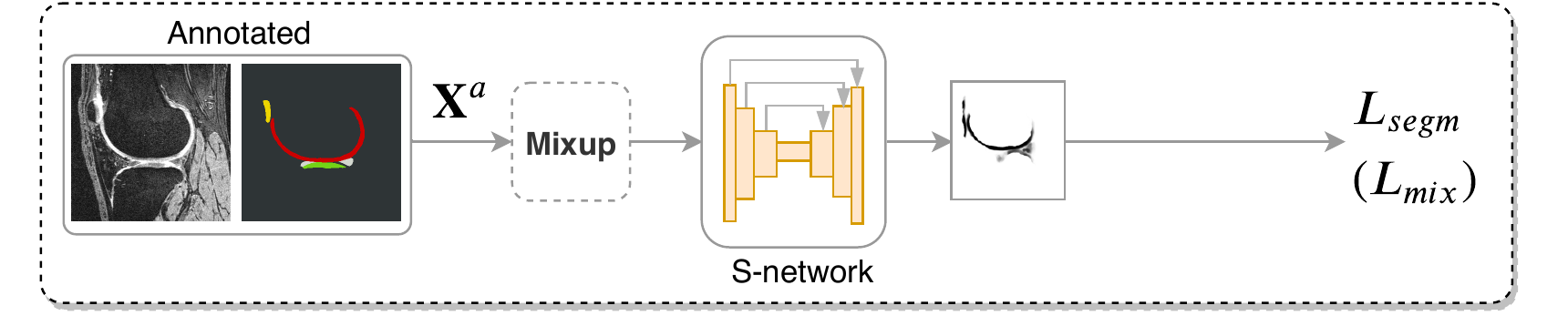}}
    \hfill
    \subfloat[\label{fig:scheme_da}]{
      \includegraphics[trim=12 5 14 0,clip,width=1.0\linewidth]{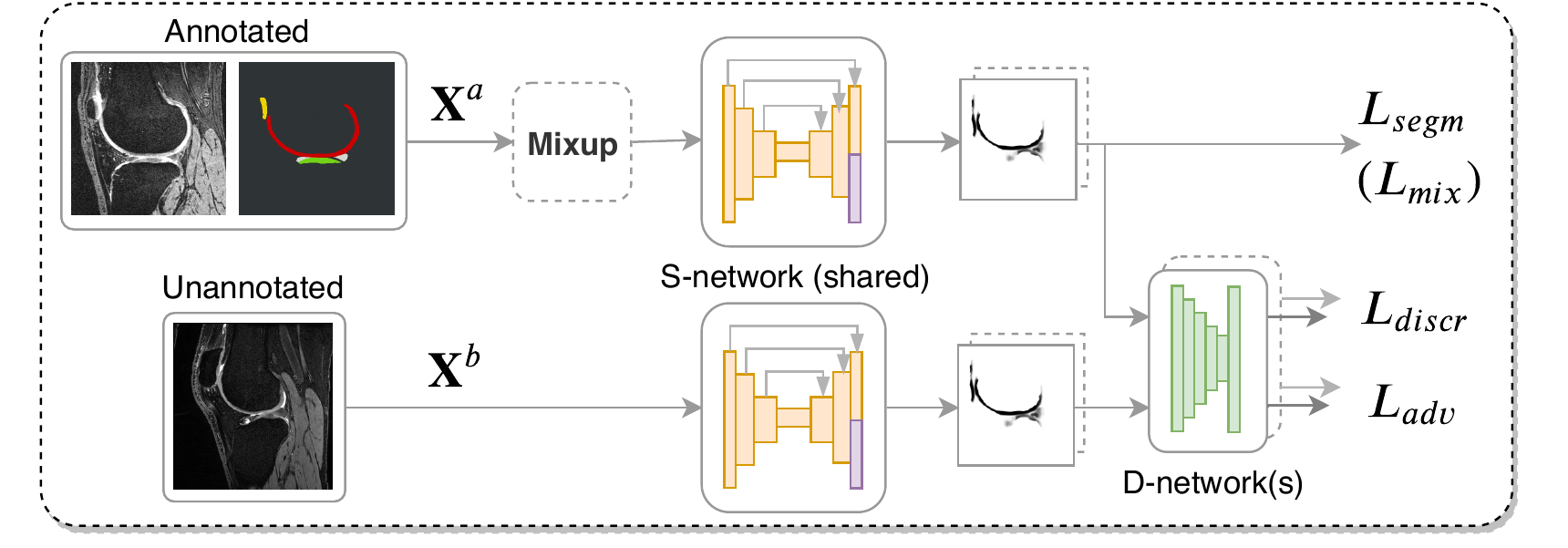}}
    \caption{Schematic view of our approaches -- without \protect\subref{fig:scheme_bl} and with \protect\subref{fig:scheme_da} UDA. Mixup, if used, is applied only during the training. In UDA setting \protect\subref{fig:scheme_da}, the S- and D- networks are trained in an adversarial manner. During the testing, only the S-network is utilized.}
    \end{center}
    \label{fig:scheme_overview}
\end{figure}
\paragraph{Mixup.}
We followed the original implementation\footnote{\url{https://github.com/facebookresearch/mixup-cifar10}} and adapted mixup to our problem. Here, the samples from the mini-batch and its permuted version were paired. Then, the virtual inputs were constructed from the pairs and passed through the network $S$:
\begin{align}
    \lambda &\sim \operatorname{Beta}(\alpha, \alpha) \label{eq:mixup_lambda} \\
    \textbf{X}_{perm}, \textbf{Y}_{perm} &= \operatorname{permute(\textbf{X}, \textbf{Y})} \\
    \textbf{X}_{mix} &= \lambda \textbf{X} + (1 - \lambda) \textbf{X}_{perm} \label{eq:mixup_x} \\
    \hat{\textbf{Y}} &= S(\textbf{X}_{mix}) \label{eq:mixup_pass} ,
\end{align}
where parameter $\alpha$ configured the augmentation strength.
The segmentation loss for the settings with mixup was defined using the model predictions and the corresponding pairs of reference annotations:
\begin{equation}\label{eq:loss_mixup}
    L_{mix} = \lambda L_{segm}(\hat{\textbf{Y}}, \textbf{Y}) + (1 - \lambda) L_{segm}(\hat{\textbf{Y}}, \textbf{Y}_{perm})
\end{equation}
\paragraph{Unsupervised Domain Adaptation.}
Differently from mixup, UDA allows to utilize both labelled and unlabelled data. In this study we adapted the method from~\cite{tsai2018structured} (see Figure~\ref{fig:scheme_da}). In particular, the segmentation model here was trained to produce the representations that do not incorporate dataset-specific biases. This was done by aligning the output and, optionally, penultimate representation spaces of the network across the datasets. For this, two networks -- a segmentation network $S$ and a discriminator network $D$ -- were trained in an adversarial manner.
Network $S$ was trained to predict the segmentation masks of the cartilage tissues and menisci by minimizing a sum of losses:
\begin{equation}\label{eq:sn_criterion}
\begin{split}
    \gamma_{segm} L_{segm}(\mathbf{S}, \mathbf{X}^a, \mathbf{Y}^a) & + \\ \gamma_{adv} L_{adv}(\mathbf{D}, \mathbf{S}, \mathbf{X}^b) & \rightarrow \min_{\mathbf{S}},
\end{split}
\end{equation}
where $\mathbf{S}$ and $\mathbf{D}$ are parameters of $S$ and $D$, and $L_{adv}$ is an adversarial loss:
\begin{equation}\label{eq:loss_adv}
    L_{adv}(\mathbf{D}, \mathbf{S}, \mathbf{X}^b) = \textrm{BCE}(\textbf{0}, D(S(\mathbf{X}^b))),
\end{equation}
where $\textrm{BCE}$ is the binary cross-entropy and $\textbf{0}$ is a matrix of all zeros having the same shape as $\mathbf{X}^b$.

Network $D$, which enforced the domain-agnostic behaviour of $S$, acted as a domain discriminator and was trained as follows:
\begin{equation} \label{eq:loss_discr_sum}
    L_{discr}(\textbf{D}, \textbf{S}, \mathbf{X}^a, \mathbf{X}^b) = L_{discr}^0 + L_{discr}^1 \rightarrow \min_{\mathbf{D}},
\end{equation}
where
\begin{align}
    L_{discr}^0(\textbf{D}, \textbf{S}, \mathbf{X}^a) &= \textrm{BCE}(\textbf{0}, D(S(\mathbf{X}^a))) \\
    L_{discr}^1(\textbf{D}, \textbf{S}, \mathbf{X}^b) &= \textrm{BCE}(\textbf{1}, D(S(\mathbf{X}^b)))
\end{align}
Here, $\textbf{1}$ is a matrix of all ones having the same shape as $\mathbf{X}^a$ and $\mathbf{X}^b$.
The discriminator was built from $5$ convolutional layers (with $64$, $128$, $256$, $512$, and $1$ filters) alternated by $4$ LeakyReLU and followed by bilinear upsampling to the input image shape. Hereinafter we call the described approach UDA1.

Additionally, we evaluated the extension of the method, where the adaptation is also applied to the penultimate decoder block of $S$. We hypothesized that adaptation at two levels can yield better alignment of the representations and compensate for the potential spatial shift between the domain in the output space. An ASPP block~\cite{chen2017deeplab} was added on top of the last decoder block, and its output was bilinearly upscaled to the dimensions of the $S$ output. The upscaled activations were used to compute the auxiliary segmentation loss and also as an input to the second discriminator. This discriminator had the same architecture as the first one, was trained following the same procedure, and contributed to the minimization criteria in Equation~\ref{eq:sn_criterion} with a smaller $\gamma_{adv}$. Further implementation details can be found in~\cite[Section~4.2]{tsai2018structured}. This approach with the adaptation of two representation spaces is further referred to as UDA2.
\subsection{Evaluation}\label{sec:methods_evaluation}
To assess the segmentation results, we used both planar (slice-wise) and volumetric (scan-wise) Dice similarity coefficients (DSCs). To examine the localization of the segmentation errors, we registered the scans such that the lateral and medial sides of the knees are oriented identically, mirroring the scans where needed. Subsequently, we computed the distribution of planar DSCs for each sagittal slice index over all registered scans in the test sets. The $95\%$ confidence intervals were estimated using bootstrapping.
\subsection{Data}
\begin{figure}[ht!]
    \begin{center}
    \subfloat[\label{fig:mri_scan}]{
      \includegraphics[width=.45\linewidth]{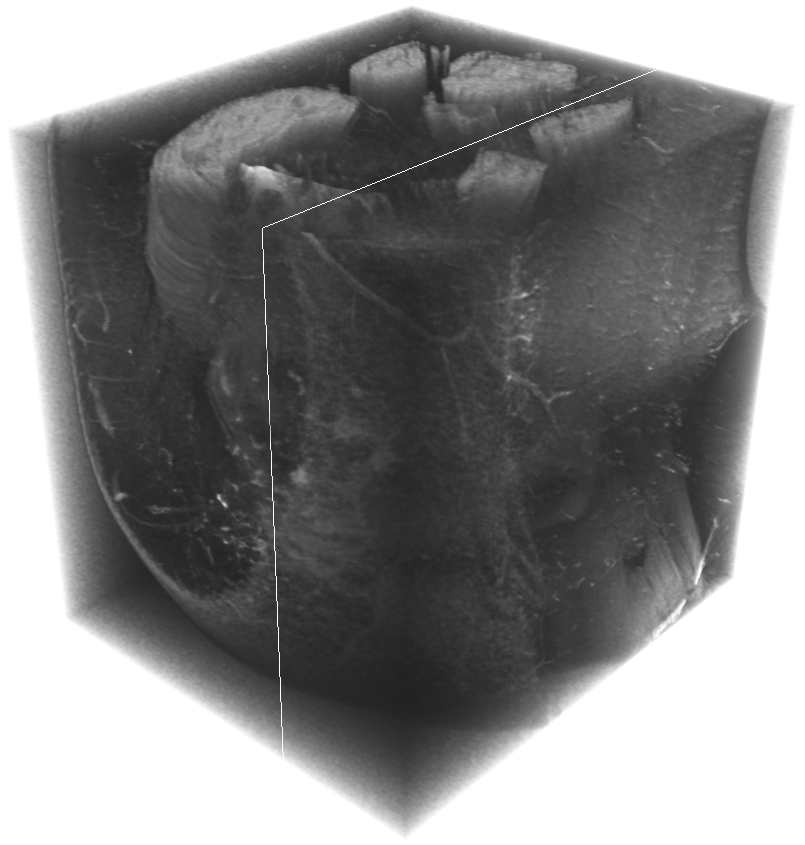}}
    \subfloat[\label{fig:mri_mask}]{
      \includegraphics[width=.45\linewidth]{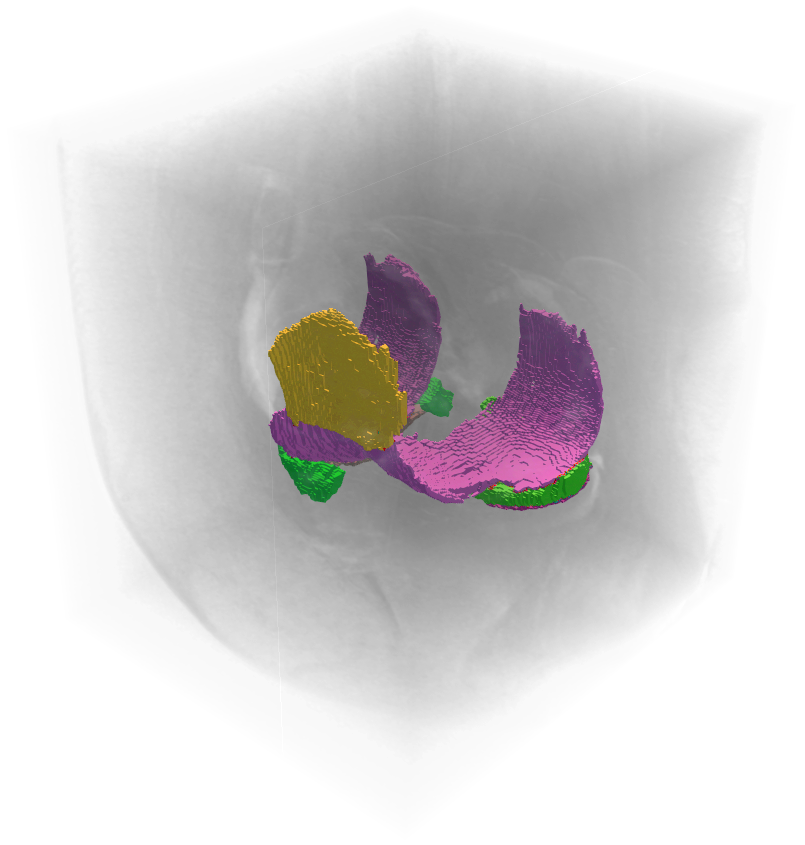}}
    \caption{DESS MRI scan \protect\subref{fig:mri_scan} and the annotations of knee cartilage and meniscal tissues \protect\subref{fig:mri_mask}, both rescaled to isotropic resolution. White lines in \protect\subref{fig:mri_scan} indicate the orientation of sagittal slices.}
    \end{center}
    \label{fig:mri}
\end{figure}

\paragraph{Overview.}
As mentioned previously, our setup included three different datasets of knee sagittal 3D double echo steady state (DESS) MRI (see Figure~\ref{fig:mri_scan}): Datasets A, B, and C. Datasets B and C were collected at our hospital using the same scanner. Dataset A contained the data from a different scanner and a distinct imaging protocol (see examples in Figure~\ref{fig:datasets}).

All the datasets were supplemented with the Kellgren-Lawrence (KL) scores~\cite{kellgren1957radiological}, derived from the radiographs associated to each subject. KL grading is the gold standard system for radiographic OA severity assessment. According to this scale, OA severity is graded from KL$0$ (no visible OA) to KL$4$ (end stage OA).
\paragraph{Dataset A.} 
The data were obtained from the Osteoarthritis Initiative (OAI, \url{http://www.oai.ucsf.edu/}) database. Dataset A included $88$ subjects from the OAI baseline and $12$-month follow-up examinations: $0$/$4$/$59$/$101$/$12$ scans with KL$0$/$1$/$2$/$3$/$4$ respectively. The data were acquired with 3T Siemens MAGNETOM Trio scanners and quadrature transmit-receive knee coils (USA Instruments, Aurora, OH, USA). Sagittal DESS sequence was used ($160$ slices; voxel size: $0.37 \times 0.37 \times 0.7mm$, matrix: $384 \times 384$, field of view (FOV): $140mm$; repetition time (TR): $16.3ms$, echo time (TE): $4.7ms$, flip angle: $25^\circ$).
Manual annotations were available for femoral, tibial, and patellar cartilage tissues, and also menisci (Figure~\ref{fig:mri_mask}).

\paragraph{Dataset B.} 
The dataset included $108$ subjects: $14$/$42$/$28$/$22$/$2$ scans with KL$0$/$1$/$2$/$3$/$4$ respectively \anonymized{(ClinicalTrials.gov Identifier: NCT02937064)}. The knees were imaged with 3T Siemens MAGNETOM Skyra scanner using a $15$-channel transmit-receive knee coil (QED, Mayfield Village, OH, USA). Sagittal DESS sequence was used ($160$ slices; voxel size: $0.59 \times 0.59 \times 0.6mm$, matrix: $256 \times 256$, FOV: $150mm$; TR: $14.1ms$, TE: $5ms$, flip angle: $25^\circ$).
No reference annotations for the tissues were available.
\paragraph{Dataset C.} 
The dataset~\anonymized{\cite{podlipska2016comparison}} included $44$ subjects: $0$/$16$/$13$/$15$/$0$ scans with KL$0$/$1$/$2$/$3$/$4$ respectively. The scanner and the data acquisition protocol were the same as for Dataset B. The annotations were produced by the research group and consisted of the segmentation masks for femoral and tibial cartilage tissues.
\subsection{Implementation details}
\paragraph{Data Pre-processing.}
In our experiments, firstly, all the data were rescaled to the pixel size of $0.37 \times 0.37$ mm. Secondly, the intensity histograms of the images were truncated (from $10^{th}$ to $99^{th}$ percentiles). Finally, the central image regions of $300 \times 300$ pixels were cropped and used for training and evaluation.

To avoid overfitting, we used the data augmentations during the training. In particular, we used random left-right flipping, gamma correction, randomly applied downscaling followed by upscaling, and also bilateral filtering.
\paragraph{Training.}
We trained and evaluated the described regularization techniques in several settings. The baseline segmentation network was trained from scratch as-is, with mixup (with and without weight decay), with UDA1, with UDA1 and mixup (without weight decay for $S$), and with UDA2.

Dataset A was randomly split into the train and the test subsets using stratification by subject ID and balancing with respect to KL grading scores. Dataset B was used solely for training and validation, Dataset C -- solely for testing. All the methods were trained using $5$-fold cross-validation following the stratification strategy described above. For the methods with UDA, Dataset B was similarly divided into $5$ folds and randomly combined with the folds of Dataset A. Therefore, for each of the experiment we train $5$ models. During the testing the predictions of these models were averaged.

In all of the experiments we used Adam~\cite{kingma2014adam}, one independent optimizer per network, depending on the setting. For all experiments with weight decay, the regularization constant was set to $5 \cdot 10^{-5}$. Parameter $\alpha$ in mixup, which is, typically, in the range from $0$ to $1$, was set to $0.7$.

The baseline method with and without mixup was trained for $50$ epochs starting with the learning rate (LR) of $10^{-3}$ (reduced to $10^{-4}$ at the $30^{th}$ epoch). All the variants with domain adaptation were trained for $30$ epochs. Initial learning rates were set to $10^{-4}$ for $S$ and $4 \times 10^{-5}$ for $D$, both reduced by a factor of $10$ at the $25^{th}$ epoch. $S$ and $D$ were updated alternately on each batch of images. For the experiments with mixup, Equation~\ref{eq:loss_mixup} was used as a segmentation loss instead of $\textrm{MCE}$ loss. In UDA1 experiments $\gamma_{segm}$ and $\gamma_{adv}$ were set to $1$ and $10^{-3}$ respectively in order to prioritize the segmentation task in the adversarial training. In UDA2 experiment, in addition to the above, auxiliary $\gamma_{segm}$ and $\gamma_{adv}$ were set to $10^{-1}$ and $2 \times 10^{-4}$ respectively, following the original publication. In the experiment with combination of mixup and UDA approaches, we applied mixup only for segmentation task. Additional forward pass of $S$ with unmixed data and without accumulation of gradients was performed to produce the input for $D$. Otherwise, the UDA architecture was kept the same.

Testing of the methods was performed on the test subset of Dataset A and whole Dataset C. To conduct the experiments we used NVIDIA 2080 Ti GPU and PyTorch~\cite{paszke2017pytorch}.

\section{Results}
\begin{table*}[ht!]
    \footnotesize
    \begin{center}
    \begin{tabular}{l|c|cc|c|cc}
    \toprule
    \multicolumn{1}{c|}{\multirow{2}{*}{\textbf{Method}}} & \textbf{FC} & \multicolumn{2}{c|}{\textbf{TC}} & \textbf{PC} & \multicolumn{2}{c}{\textbf{M}} \\
    \multicolumn{1}{c|}{} &  & \textit{medial} & \textit{lateral} &  & \textit{medial} & \textit{lateral} \\
    \midrule
    Norman et al.~\cite{norman2018relaxometry} & 0.867\ws(0.032) & 0.777\ws(0.029) & 0.799\ws(0.036) & 0.767\ws(0.091) & 0.731\ws(0.054) & 0.812\ws(0.030) \\
    Tack et al.~\cite{tack2018menisci} & - & - & - & - & 0.838\ws(0.061) & 0.889\ws(0.024) \\
    Ambellan et al.~\cite{ambellan2019bonecartilage} & 0.894\ws(0.024) & 0.861\ws(0.053) & 0.904\ws(0.024) & - & - & - \\
    Desai et al.~\cite{desai2019considerations} & 0.89\hphantom{d}\ws(0.02\hphantom{d}) & - & - & - & - & - \\
    Tack et al.~\cite{tack2019accurate} & - & 0.880\ws(0.046) & 0.913\ws(0.023) & - & - & - \\
    Chaudhari et al.~\cite{chaudhari2019utility} & 0.902\ws(0.017) & - & - & - & - & - \\
    Ours (baseline method) & 0.907\ws(0.019) & \multicolumn{2}{c|}{0.897\ws(0.028)} & 0.871\ws(0.046) & \multicolumn{2}{c}{0.863\ws(0.034)} \\
    \bottomrule
    \end{tabular}%
    \end{center}
    \caption{Comparison of our baseline to the previously published methods on Dataset A. Numbers are the means and standard deviations of volumetric DSCs. The scores are given for reference and should be compared carefully.~\cite{norman2018relaxometry,desai2019considerations,chaudhari2019utility} used slightly different train/validation/test splits.~\cite{chaudhari2019utility} performed segmentation in 3D.~\cite{tack2018menisci,ambellan2019bonecartilage,tack2019accurate} used multi-stage pipelines (2D segmentation, statistical shape modelling, 3D refinement), $2$-fold cross validation, and reported the results for $2$ examinations (we present only the highest scores).}
    \label{table:scores_base}
\end{table*}
\paragraph{Baseline.}
We compared our baseline method to the published state-of-the-art approaches in knee cartilage and menisci segmentation (see Table~\ref{table:scores_base}). Our baseline method performed either more accurately or on par with others depending on the tissue. Even though it was not designed to separate lateral and medial tibial cartilage tissues and menisci as in~\cite{tack2018menisci,ambellan2019bonecartilage,tack2019accurate}, it provided other advantages, namely, it was faster in training and inference, more lightweight, and produced masks for all the considered tissues simultaneously.

\begin{figure*}[ht!]
    \footnotesize
    \begin{center}
        \begin{tabular}
            {m{0.07\textwidth}@{\hspace{3mm}} *3{>{\hspace{0mm}\centering\arraybackslash}m{0.30\textwidth}@{\hspace{1mm}}}}
        & {Dataset A (end-stage OA)} & {Dataset C (healthy)} & {Dataset C (severe OA)} \\
        {Input} &
        \includegraphics[width=.30\textwidth]{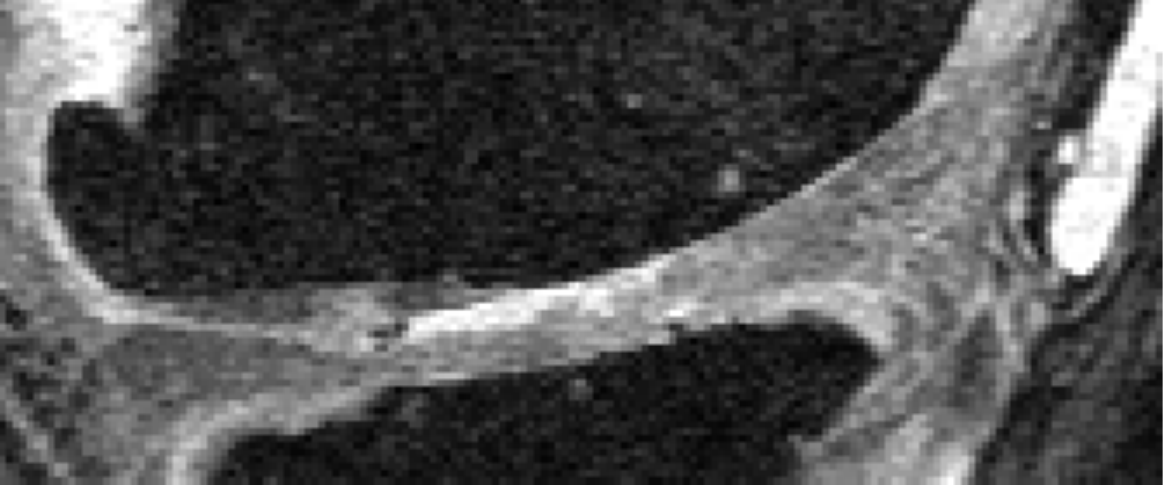} &
        \includegraphics[width=.30\textwidth]{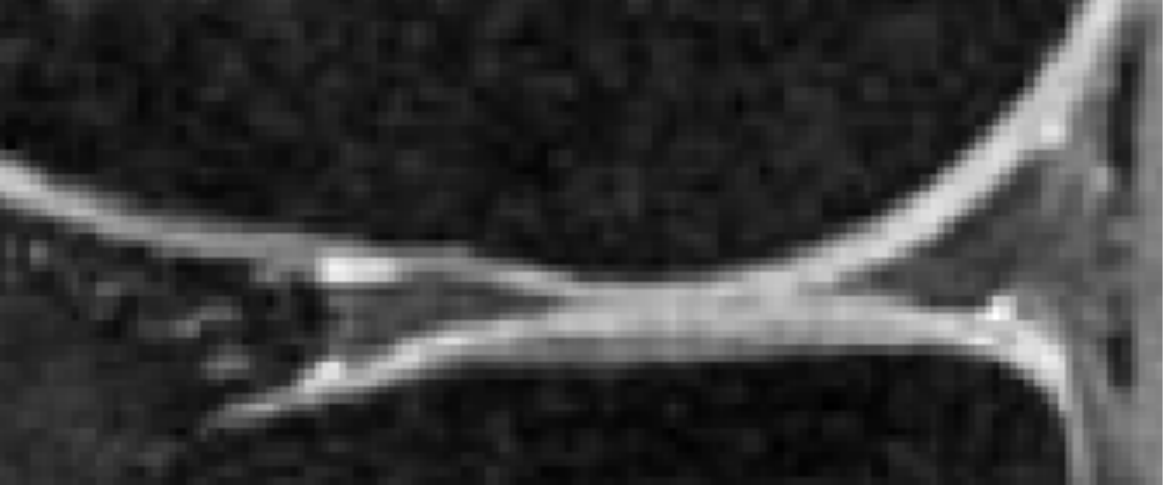} &
        \includegraphics[width=.30\textwidth]{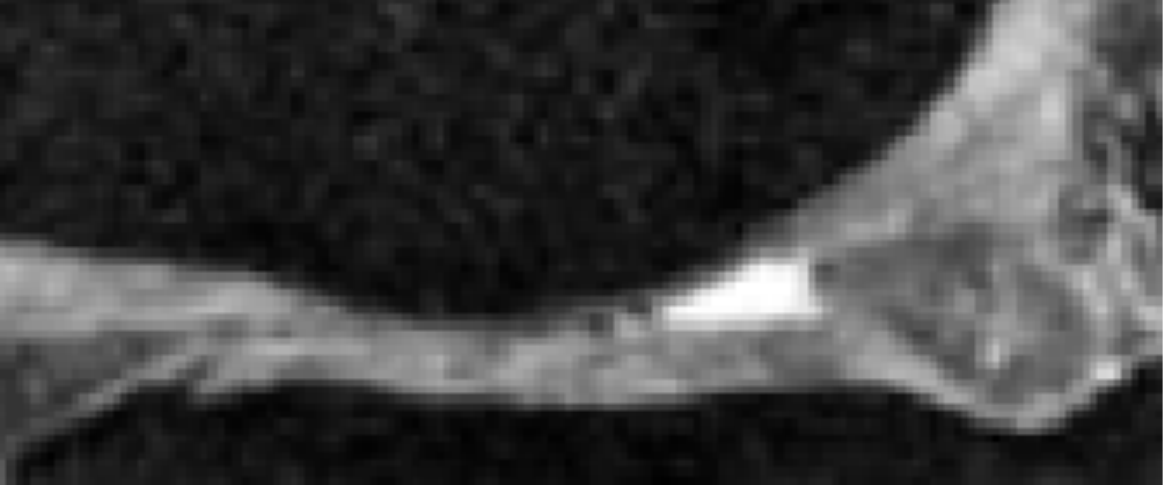} \\
        {Ground \newline truth} &
        \includegraphics[width=.30\textwidth]{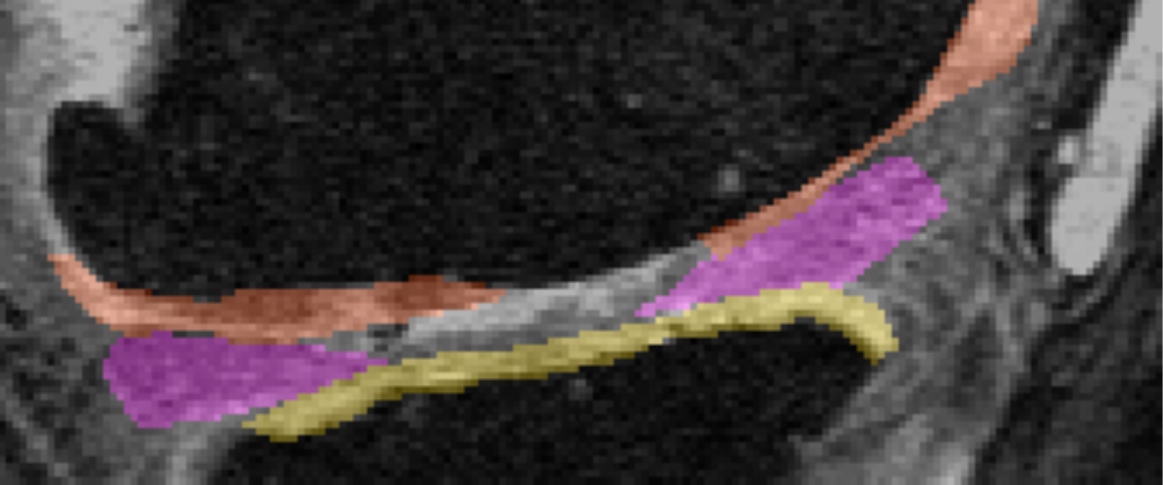} &
        \includegraphics[width=.30\textwidth]{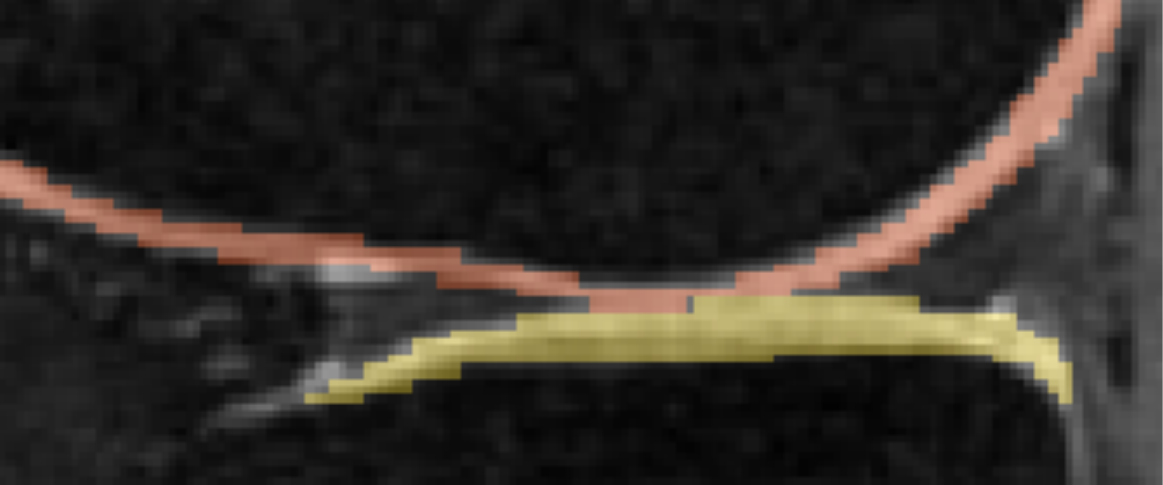} &
        \includegraphics[width=.30\textwidth]{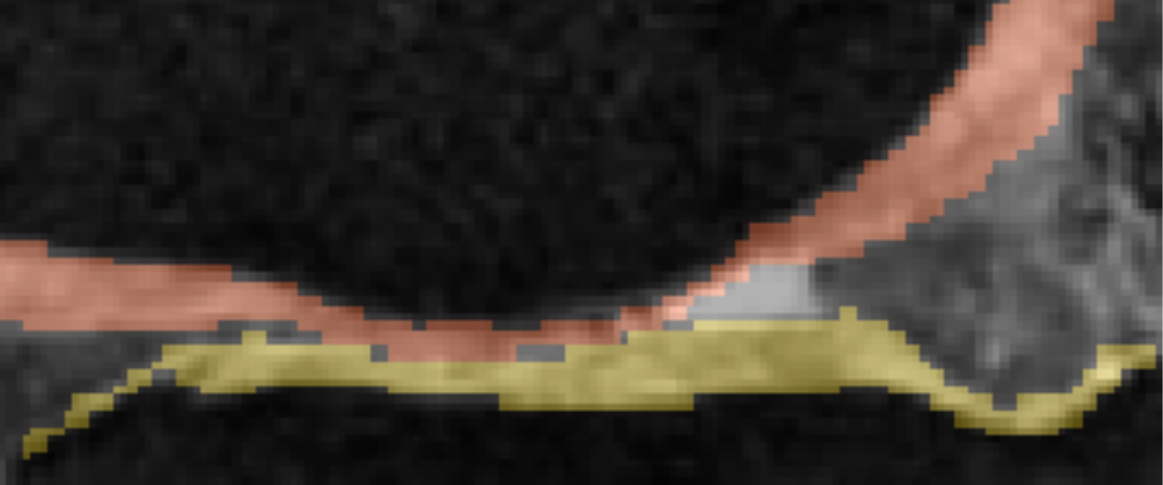} \\
        {Baseline} &
        \includegraphics[width=.30\textwidth]{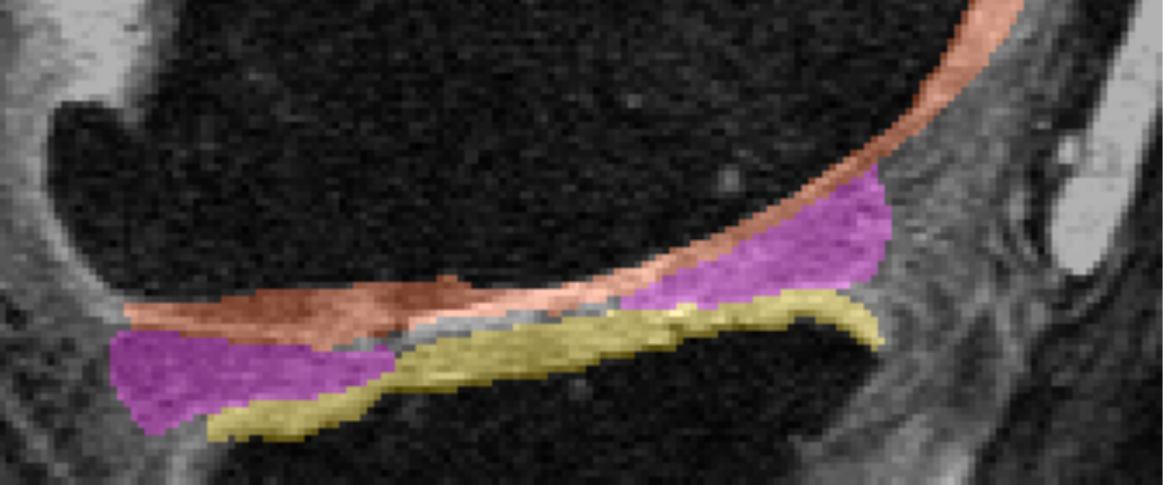} &
        \includegraphics[width=.30\textwidth]{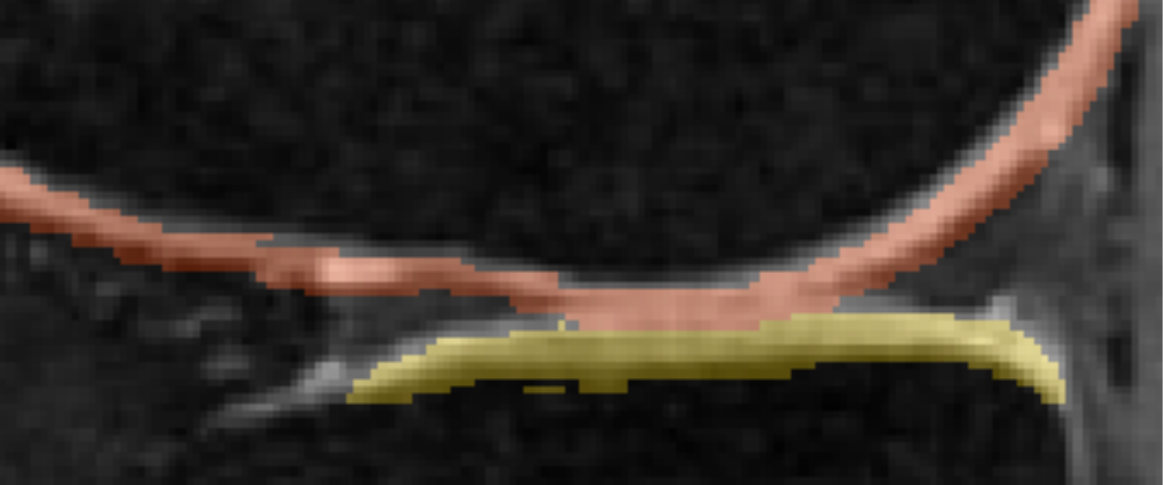} &
        \includegraphics[width=.30\textwidth]{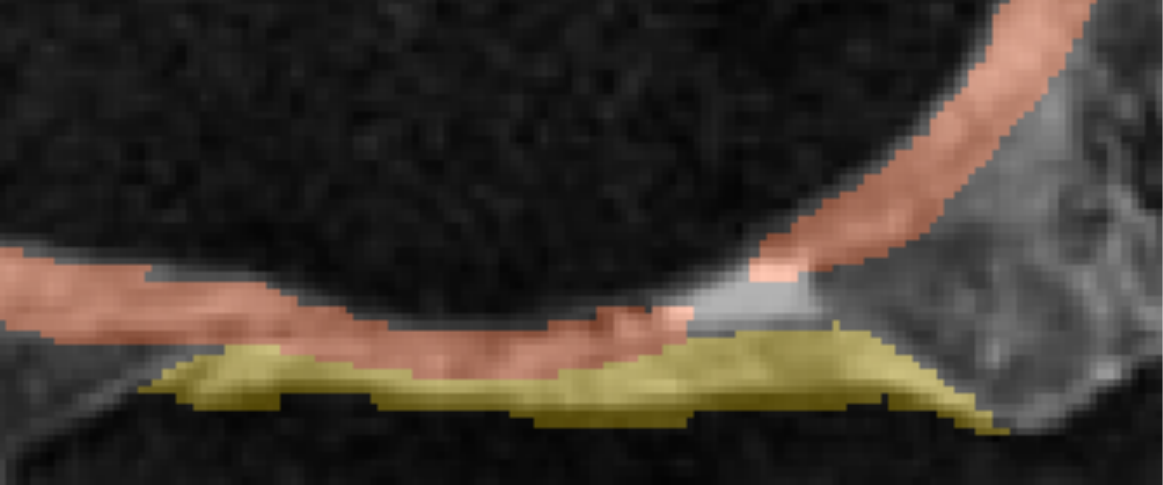} \\
        {+ mixup \newline - WD} &
        \includegraphics[width=.30\textwidth]{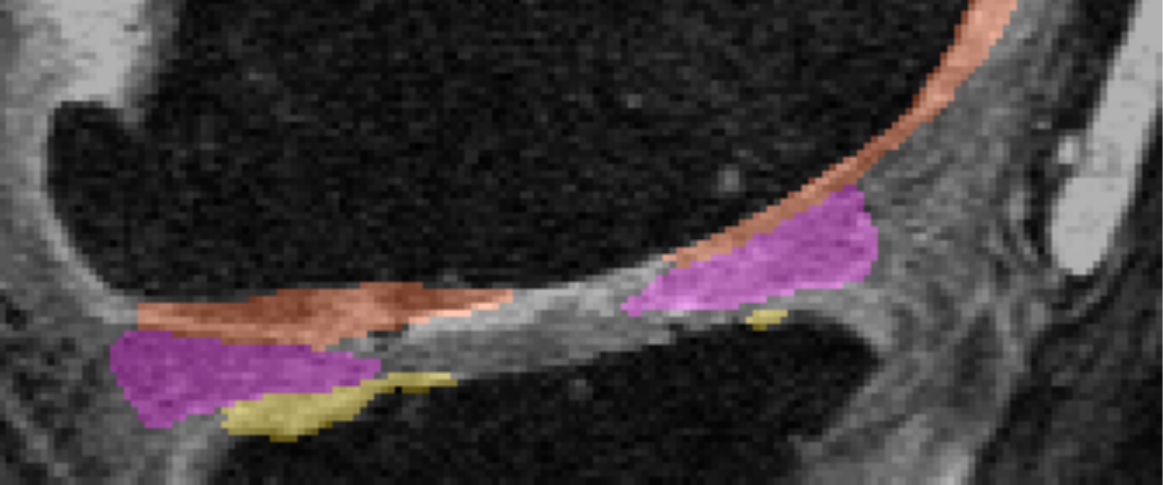} &
        \includegraphics[width=.30\textwidth]{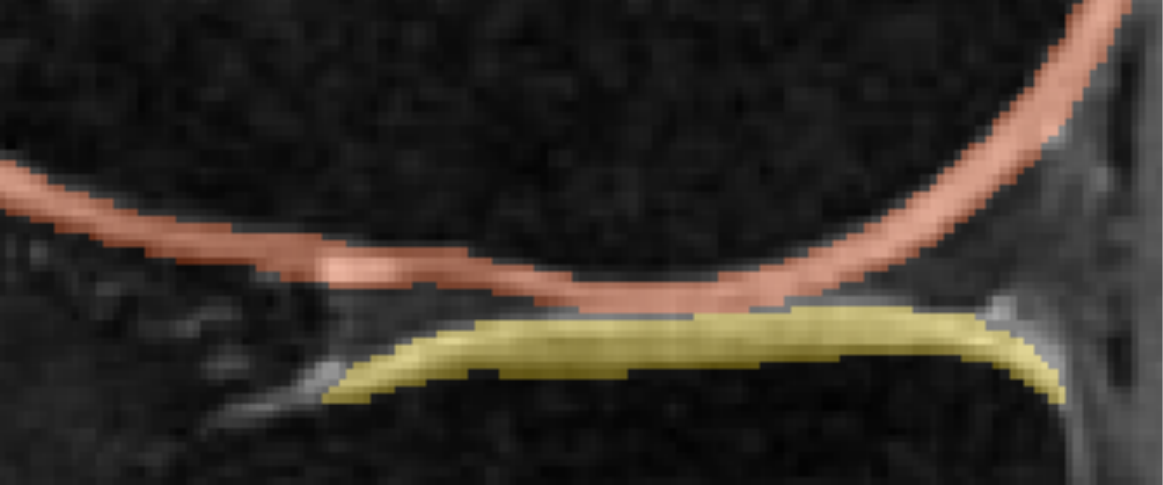} &
        \includegraphics[width=.30\textwidth]{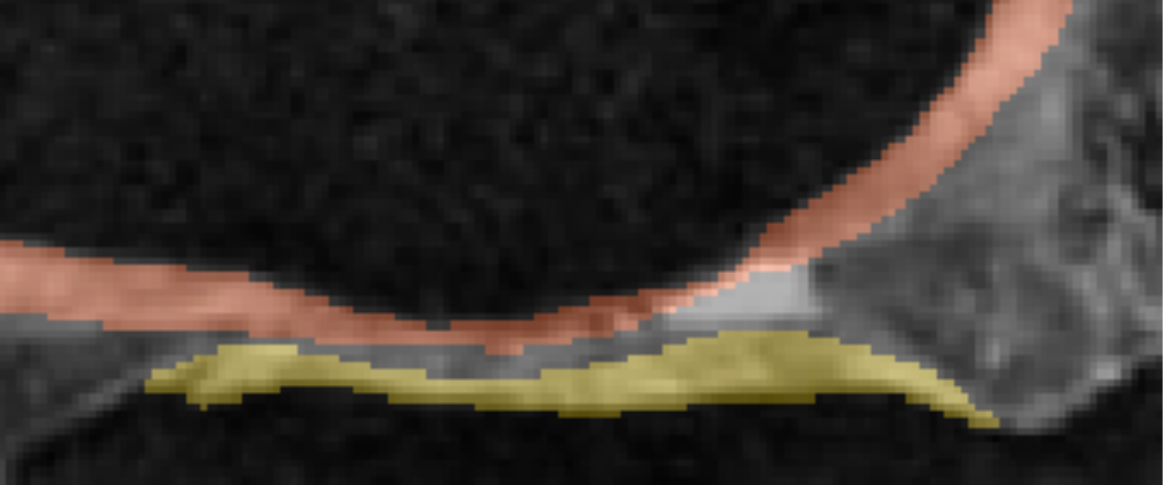} \\
        {+ UDA2} &
        \includegraphics[width=.30\textwidth]{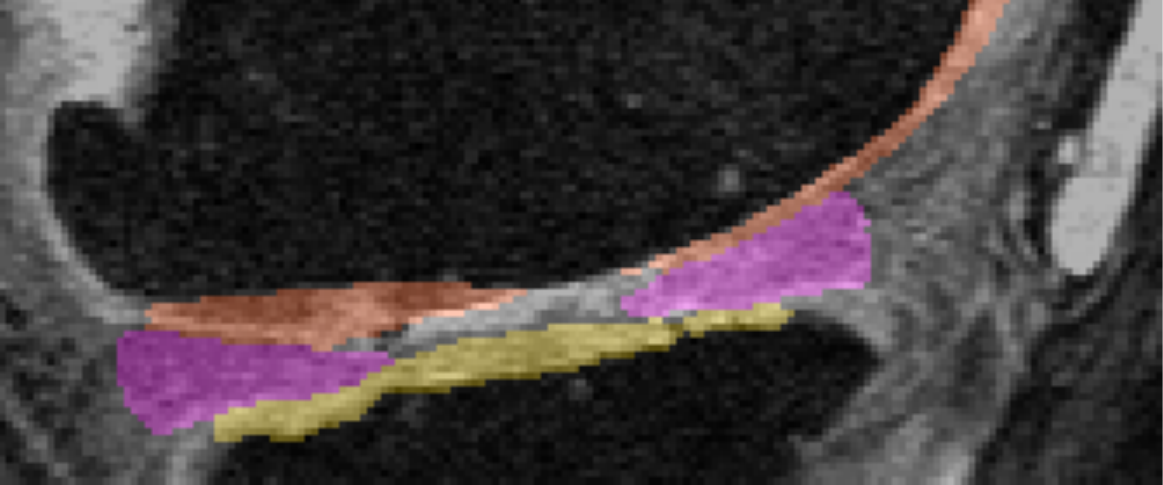} &
        \includegraphics[width=.30\textwidth]{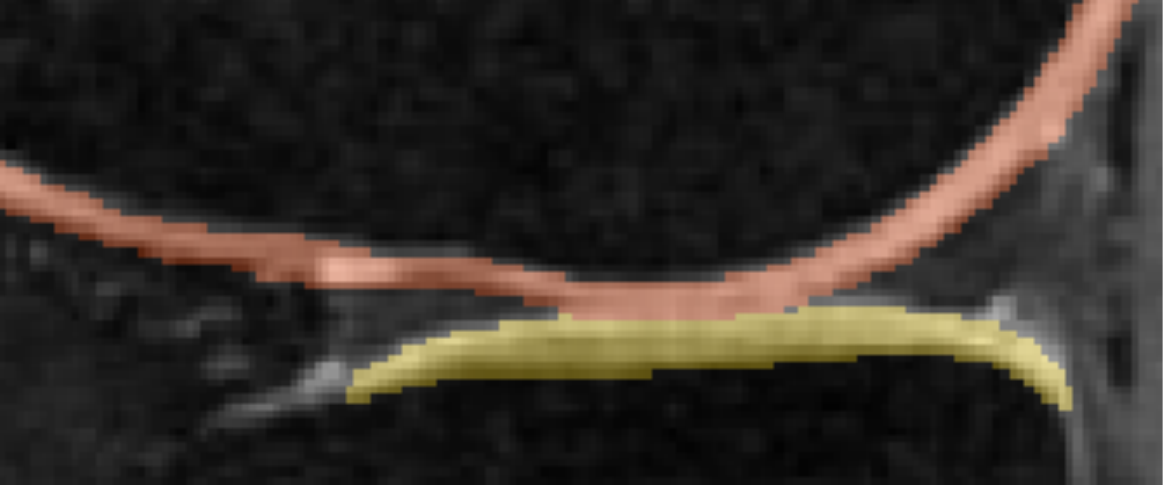} &
        \includegraphics[width=.30\textwidth]{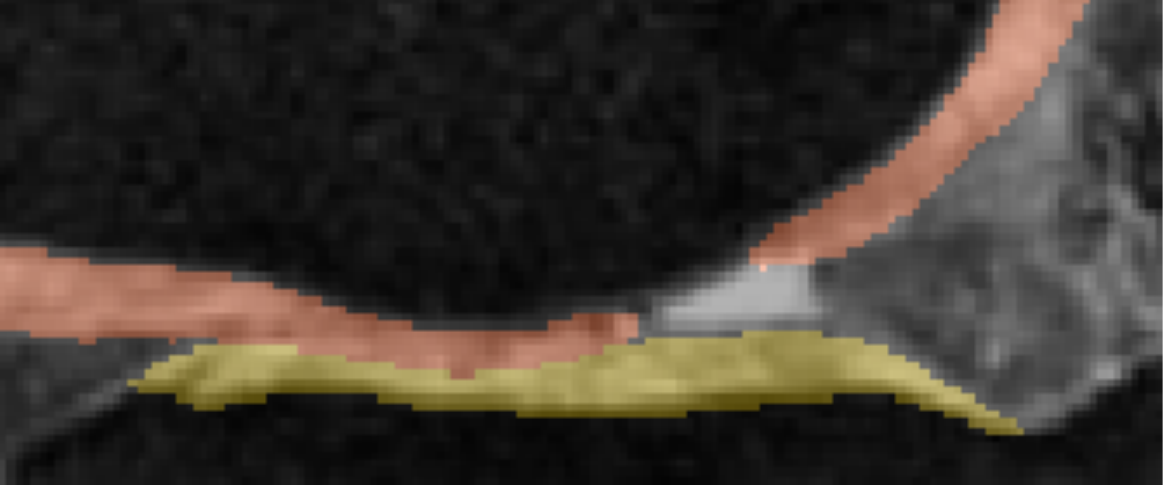} \\
        \end{tabular}
    \end{center}
    \caption{Example images of tibiofemoral contact zones from Datasets A and C, respective annotations, and the segmentation masks produced by the baseline and the regularized approaches. Visual differences between the datasets can be observed. Colors highlight cartilage tissues: orange -- femoral, yellow -- tibial, purple -- menisci. Patellar cartilage was not presented in the considered Dataset A slice, for that reason patellofemoral zone is not shown.}
    \label{fig:results_comparison_main}
\end{figure*}

On Dataset A the method reached $0.907\, (0.019)$ for FC and $0.897\, (0.028)$ for TC, however, on Dataset C the scores for the respective tissues were $0.791\, (0.033)$ and $0.746\, (0.037)$. Such discrepancy in the scores was, presumably, caused by the several factors: lack of model robustness, which resulted in biased and noisy segmentations (see Figure~\ref{fig:results_comparison_main}), and lower original resolution of images and annotations in Dataset C, which made the segmentation more challenging and increased the cost of annotation errors.

\paragraph{Mixup.}
We found that applying mixup lead to a minor underfitting on Dataset A (see Table~\ref{table:scores_main}), yet the generalization had increased ($0.804\, (0.031)$ for FC, $0.791\, (0.034)$ for TC on Dataset C). Such phenomena was also reported for object classification in Verma~\emph{et al.}~\cite{verma2018manifold}. However, since mixup itself is a strong regularizer, we hypothesized that avoiding the use of weight decay could address the undefitting. In this new setting our model largely recovered the scores on Dataset A and further improved the performance for FC on Dataset C ($0.819\, (0.025)$ for FC, $0.802\, (0.029)$ for TC).
\paragraph{Unsupervised Domain Adaptation.}
An approach with UDA1 on Dataset C yielded comparable DSCs ($0.815\, (0.025)$ for FC, $0.814\, (0.029)$ for TC) to the best mixup setting, however, the scores on Dataset A were lower (see Table~\ref{table:scores_main}). Performing representation alignment at the multiple layers of the network (UDA2) improved on top of UDA1 for Dataset A. However, on Dataset C the performance increased only for FC, while it became worse for TC ($0.821\, (0.025)$ for FC, $0.799\, (0.033)$ for TC). What concerns the efficiency, the computational costs for training UDA approaches were up to three times higher compared to mixup.
\begin{table*}[]
    \footnotesize
    \begin{center}
    \begin{tabular}{l|cccc|cc}
    \toprule
    \multicolumn{1}{c}{\multirow{2}{*}{\textbf{Method}}} & \multicolumn{4}{c}{\textbf{Dataset A}} & \multicolumn{2}{c}{\textbf{Dataset C}} \\
     & \textbf{FC} & \textbf{TC} & \textbf{PC} & \textbf{M} & \textbf{FC} & \textbf{TC} \\
    \midrule
    Baseline & \textbf{0.907\ws(0.019)} & \textbf{0.897\ws(0.028)} & \textbf{0.871\ws(0.046)} & \textbf{0.863\ws(0.034)} & 0.791\ws(0.033) & 0.746\ws(0.037) \\
    + mixup & 0.903\ws(0.019) & 0.892\ws(0.031) & \underline{0.865\ws(0.054)} & 0.852\ws(0.035) & 0.804\ws(0.031) & 0.791\ws(0.034) \\
    + mixup - WD & \textbf{0.907\ws(0.019)} & \underline{0.896\ws(0.028)} & 0.864\ws(0.054) & \underline{0.861\ws(0.033)} & \underline{0.819\ws(0.025)} & 0.802\ws(0.029) \\
    + UDA1 & 0.896\ws(0.023) & 0.887\ws(0.031) & 0.852\ws(0.064) & 0.851\ws(0.035) & 0.815\ws(0.025) & \textbf{0.814\ws(0.029)} \\
    + UDA2 & 0.901\ws(0.021) & 0.892\ws(0.031) & 0.861\ws(0.060) & 0.856\ws(0.035) & \textbf{0.821\ws(0.025)} & 0.799\ws(0.033) \\
    + mixup - WD + UDA1 & 0.895\ws(0.023) & 0.886\ws(0.027) & 0.846\ws(0.066) & 0.849\ws(0.034) & 0.810\ws(0.026) & \underline{0.810\ws(0.031)} \\
    \bottomrule
    \end{tabular}
    \end{center}
    \caption{Regularization approaches evaluated tissue-wise on two datasets. Numbers are the means and standard deviations of volumetric DSCs. The best score for each tissue is highlighted in bold, the second best - is underlined. "- WD" indicates the experiments without weight decay. }
    \label{table:scores_main}
\end{table*}
\paragraph{Combined Approach.}
We experimented with combining mixup and UDA approaches. Here, we took UDA1 setting and applied mixup to the supervised segmentation task. Otherwise, the architecture was kept the same, including the input to the generator. The approach with both mixup and UDA1 showed the worst performance among others on Dataset A, yet still showed better DSCs on Dataset C compared to the baseline.
\paragraph{Detailed Analysis.}
As previously said, cartilage tissues degenerate over the progression of OA. Lesions start to appear, cartilage is getting worn out and, therefore, it becomes challenging to segment. To evaluate the performance of the methods in relation to OA severity (from doubtful OA to end-stage OA), we computed the volumetric DSCs over the test sets KL-grade-wise. Here, for the sake of brevity, we compared only the baseline approach, the best of mixup, and the best of UDA. The results are presented in Table~\ref{table:scores_detailed}.

\begin{figure}[t]
    \footnotesize
    \begin{center}
    \subfloat[\label{fig:results_dsc_distr_a}]{
        \includegraphics[trim=7 7 7 7,clip,width=0.95\linewidth]{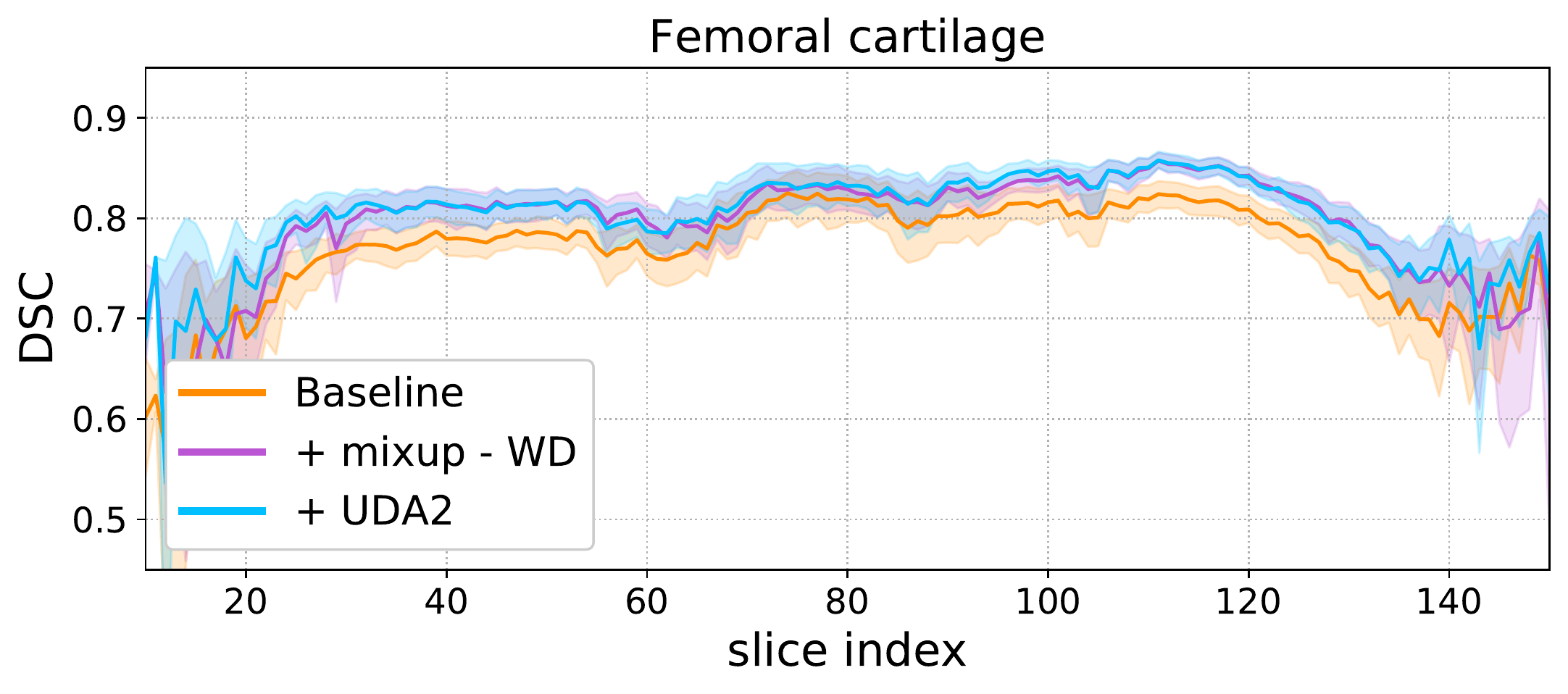}}
    \hfill\null
    \vfill
    \subfloat[\label{fig:results_dsc_distr_b}]{
        \includegraphics[trim=7 7 7 7,clip,width=0.95\linewidth]{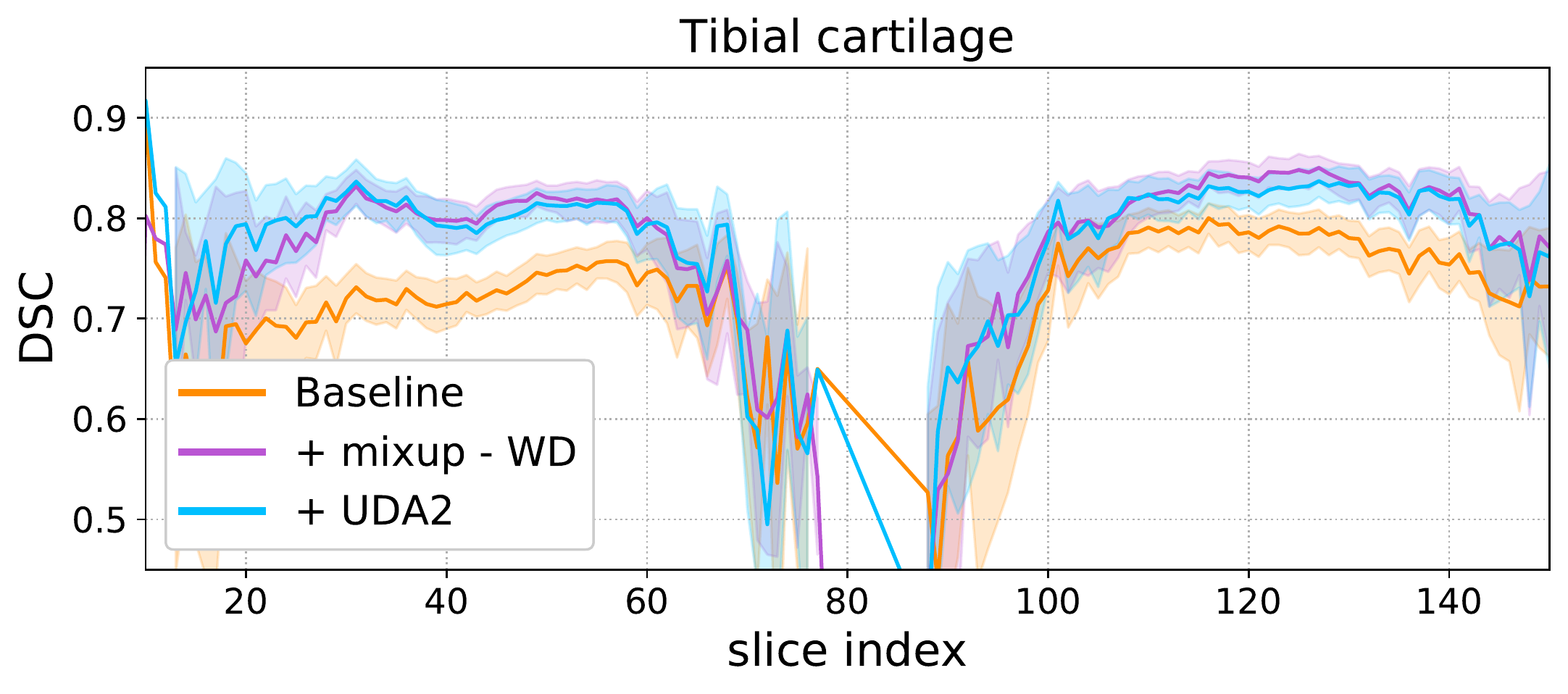}}
    \hfill\null
    \caption{Distributions of the planar DSCs computed slice-wise (from 0th to 159th slice, medial to lateral, respectively). Solid lines indicate the distribution means, bright bands -- the 95\% confidence intervals. Slices approx. 20-60 and 100-140 correspond to the locations of the medial and lateral femoral condyles (i.e. weight-bearing areas of the joint). Slices approx. 60-100 enclose the intercondylar notch and, therefore, are of less clinical interest.}
    \end{center}
    \label{fig:results_dsc_distr}
\end{figure}

The detailed analysis showed that both modifications (mixup and UDA2) yielded similar significant improvements on Dataset C for most of the cases, while the approach with mixup better maintained the performance on Dataset A. The results were further analyzed with respect to anatomical location by following the approach described in Section~\ref{sec:methods_evaluation}. Illustrated in Figure~\ref{fig:results_dsc_distr}, the results indicated that both mixup and UDA2 improved the segmentation accuracy for FC, with higher increase in the weight bearing areas of femoral condyles. For TC, the improvements were concentrated near the tibial plateaus, mainly located on the medial side.

Several challenging examples of MR images, with the corresponding annotations and the segmentation masks produced by the methods, are presented in Figure~\ref{fig:results_comparison_main}. In these cases, the baseline and UDA2 approaches tended to over-segment the tissues, while the baseline also produced shifted segmentations on Dataset C. The images also highlight inaccuracies of the reference segmentations and common limitations of the approaches.
\begin{table*}[]
    \footnotesize
    \begin{center}
    \begin{tabular}{c|c|c|cccc|c|cc}
    \toprule
    \multirow{2}{*}{\textbf{Method}} & \multirow{2}{*}{\textbf{KL}} & \multicolumn{5}{c}{\textbf{Dataset A}} & \multicolumn{3}{c}{\textbf{Dataset C}} \\
     &  & \textbf{\#} & \textbf{FC} & \textbf{TC} & \textbf{PC} & \textbf{M} & \textbf{\#} & \textbf{FC} & \textbf{TC} \\
    \midrule
    \multirow{5}{*}{Baseline} & 1 & - & - & - & - & - & 16 & 0.785\ws(0.041) & 0.757\ws(0.037) \\
     & 2 & 11 & 0.920\ws(0.015) & 0.921\ws(0.010) & 0.860\ws(0.061) & 0.873\ws(0.047) & 13 & 0.794\ws(0.031) & 0.735\ws(0.028) \\
     & 3 & 21 & 0.904\ws(0.019) & 0.891\ws(0.026) & 0.875\ws(0.040) & 0.860\ws(0.025) & 15 & 0.794\ws(0.024) & 0.743\ws(0.041) \\
     & 4 & 4 & 0.892\ws(0.003) & 0.861\ws(0.019) & 0.882\ws(0.015) & 0.854\ws(0.029) & - & - & - \\
     & all & 36 & 0.907\ws(0.019) & 0.897\ws(0.028) & 0.871\ws(0.046) & 0.863\ws(0.034) & 44 & 0.791\ws(0.033) & 0.746\ws(0.037) \\
    \midrule
    \multirow{5}{*}{\begin{tabular}[c]{@{}c@{}}+ mixup\\ - WD\end{tabular}} & 1 & - & - & - & - & - & 16 & \textbf{0.826\ws(0.024)} & \textbf{0.818\ws(0.025)} \\
     & 2 & 11 & 0.921\ws(0.013) & 0.922\ws(0.007) & 0.860\ws(0.060) & 0.872\ws(0.043) & 13 & \textbf{0.821\ws(0.023)} & \textbf{0.800\ws(0.021)} \\
     & 3 & 21 & 0.903\ws(0.019) & 0.890\ws(0.026) & \textbf{0.863\ws(0.055)} & 0.857\ws(0.026) & 15 & \textbf{0.811\ws(0.026)} & \textbf{0.787\ws(0.031)} \\
     & 4 & 4 & 0.889\ws(0.002) & 0.861\ws(0.016) & 0.877\ws(0.015) & 0.856\ws(0.027) & - & - & - \\
     & all & 36 & 0.907\ws(0.019) & 0.896\ws(0.028) & \textbf{0.864\ws(0.054)} & 0.861\ws(0.033) & 44 & \textbf{0.819\ws(0.025)} & \textbf{0.802\ws(0.029)} \\
    \midrule
    \multirow{5}{*}{+ UDA2} & 1 & - & - & - & - & - & 16 & \textbf{0.827\ws(0.024)} & \textbf{0.816\ws(0.029)} \\
     & 2 & 11 & \textbf{0.915\ws(0.015)} & 0.918\ws(0.007) & 0.851\ws(0.067) & \textbf{0.867\ws(0.045)} & 13 & \textbf{0.822\ws(0.028)} & \textbf{0.796\ws(0.022)} \\
     & 3 & 21 & \textbf{0.898\ws(0.021)} & \textbf{0.885\ws(0.030)} & \textbf{0.863\ws(0.061)} & \textbf{0.852\ws(0.027)} & 15 & \textbf{0.815\ws(0.022)} & \textbf{0.783\ws(0.036)} \\
     & 4 & 4 & 0.883\ws(0.008) & 0.856\ws(0.017) & 0.875\ws(0.019) & 0.840\ws(0.032) & - & - & - \\
     & all & 36 & \textbf{0.901\ws(0.021)} & \textbf{0.892\ws(0.031)} & \textbf{0.861\ws(0.060)} & \textbf{0.856\ws(0.035)} & 44 & \textbf{0.821\ws(0.025)} & \textbf{0.799\ws(0.033)} \\
    \bottomrule
    \end{tabular}
    \end{center}
    \caption{Comparison between the baseline and the best performing approaches. Here, means and standard deviations of volumetric DSCs are presented for the subject groups of specific KL-grades ($1$-$4$) and for the full test sets. \# shows the number of scans in the specific group. Statistically significant differences to the baseline method ($p < 0.005$ with two-sided Wilcoxon signed-rank test) are highlighted in bold. For Dataset A all the differences are either negative or insignificant, for Dataset C - either positive or insignificant.}
    \label{table:scores_detailed}
\end{table*}
\section{Conclusions}
In this study, we investigated the use of mixup and adversarial domain adaptation for DL-based knee tissue segmentation from MRI. We showed that the segmentation model trained from scratch with limited data lacked generalization and performed worse on the dataset that had different resolution and contrast. Strong regularization techniques, namely, supervised mixup and UDA, helped to partially alleviate the issue and make the model more robust. We analyzed the baseline and the best performing approaches in relation to anatomical locations and different stages of OA, and showed that the improvements over the baseline are consistent and clinically relevant.

This paper is the first to address the challenge of robustness in knee MRI segmentation in an end-to-end manner using DL. On the test set derived from OAI, our model yielded state-of-the-art segmentation results for patellar cartilage and similar to other works results for other tissues, and allowed to segment all the cartilage and meniscal tissues simultaneously.

Despite the state-of-the-art results, our study has still some limitations. In particular, we considered only 2D segmentation approach. Due to several factors, such as complex cartilage geometry, partial volume effect, lack of contextual information in 2D, and imperfect and inconsistent annotations, most of the segmentation errors produced by our methods were located on the tissue surfaces or in the slices tangential to the surfaces. Volumetric methods could potentially alleviate some of those issues and provide more accuracy and shape consistency. However, the comparison done to the previous studies~\cite{tack2019accurate,ambellan2019bonecartilage,tack2018menisci,chaudhari2019utility} showed similar performance in terms of DSCs.

Another limitation of this study is that more complex applications of mixup in UDA were not investigated. We believe that more experimental work in that direction can potentially lead to higher results. Future studies on knee MRI segmentation or medical image segmentation could further explore the potential of this idea. Besides the mentioned limitations, we acknowledge that a more comprehensive framework for assessment of regularization methods should be considered. However, in knee MRI field there is a lack of public datasets available for experiments. Our future studies will consider consolidation of various datasets in order to perform more thorough investigations, including different scanner manufacturers and diverse MRI sequences.

From the methodological point of view, our study demonstrated that for, at least, a moderate range of image variations, mixup and UDA may similarly improve the robustness of medical image segmentation. However, UDA approach is computationally heavy and difficult to train due to the need of careful hyper-parameter tuning. Besides that, our experiments showed that UDA may significantly worsen the DSCs in the source domain (Dataset A). Therefore, we think that mixup and other regularization techniques should be preferred when aiming for robust medical image segmentation using DL.

To conclude, we believe that our results will promote wider adoption of DL-based methods in OA research community and facilitate further work on development of robust segmentation methods for knee MRI. In MRI domain, such methods may become a powerful tool to leverage large and diverse imaging cohorts without available annotations and drastically speed up and improve the medical research. For instance, one important application area -- disease modifying drugs development for OA -- can directly benefit from reliable segmentations. To facilitate further knee MRI segmentation research, our source codes and pre-trained models are made publicly available:~\anonymized{\url{https://github.com/MIPT-Oulu/RobustCartilageSegmentation}}.
\section*{Acknowledgements}
The OAI is a public-private partnership comprised of five contracts
(N01-AR-2-2258; N01-AR-2-2259; N01-AR-2-2260; N01-AR-2-2261; N01-AR-2-2262)
funded by the National Institutes of Health, a branch of the Department of Health and
Human Services, and conducted by the OAI Study Investigators. Private funding partners
include Merck Research Laboratories; Novartis Pharmaceuticals Corporation,
GlaxoSmithKline; and Pfizer, Inc. Private sector funding for the OAI is managed by the
Foundation for the National Institutes of Health. This manuscript was prepared using an OAI
public use data set and does not necessarily reflect the opinions or views of the OAI
investigators, the NIH, or the private funding partners.

\anonymized{The authors would like to acknowledge the following funding sources: strategic funding of University of Oulu (Infotech Oulu), Sigrid Juselius foundation, and KAUTE foundation, Finland.}

{\small
\bibliographystyle{ieee}
\bibliography{./main.bib}
}

\end{document}